\documentclass[pre,twocolumn,amsmath,amssymb,nofootinbib,floatfix,superscriptaddress]{revtex4}

\usepackage{graphicx, bm, xcolor, bbold, enumerate, float}
\usepackage[normalem]{ulem}
\usepackage[breaklinks]{hyperref}

\makeatletter

\def\graphicscale{\twocolumn@sw{0.3}{0.4}}
\def\graphicthreescale{\twocolumn@sw{0.3}{0.4}}

\begin{document}

\title{Out-of-equilibrium percolation transitions at finite critical times \\
  after quenches across magnetic first-order transitions}

\author{Andrea Pelissetto}
\affiliation{Dipartimento di Fisica dell'Universit\`a di Roma ``La Sapienza"
  and INFN, Sezione di Roma, P.le Aldo Moro 5, I-00185 Roma, Italy}

\author{Davide Rossini} 
\affiliation{Dipartimento di Fisica dell'Universit\`a di Pisa and INFN,
  Largo Pontecorvo 3, I-56127 Pisa, Italy}

\author{Ettore Vicari} 
\affiliation{Dipartimento di Fisica dell'Universit\`a di Pisa,
  Largo Pontecorvo 3, I-56127 Pisa, Italy}

\date{\today}

\begin{abstract}
  We show that an out-of-equilibrium percolation transition occurs
  after quenching ferromagnetic Ising-like systems across their
  magnetic first-order transitions.  As a paradigmatic example, we
  consider a two-dimensional Ising system driven across its
  low-temperature first-order transition line by a quench of the
  magnetic field $h$ from $h_i<0$ to $h>0$. In the thermodynamic limit
  and for finite values of $h$, the post-quench evolution under a
  purely relaxational dynamics is characterized by a dynamic
  transition at a finite critical time $t_c(h)$ from the metastable
  negatively magnetized phase to the positive one, marked by the
  percolation of the largest clusters of positive and negative spins.
  This out-of-equilibrium percolation transition displays a
  finite-size scaling behavior as in the standard random-percolation
  case.  However, while the fractal dimension of the percolating
  clusters is consistent with the random-percolation value, the
  exponent controlling the approach to criticality differs and depends
  on $h$.  We also show that the percolation critical behavior is
  related to the spinodal-like behavior of the magnetization in the
  small-$h$ limit, which implies that the percolation time $t_c(h)$
  exhibits a spinodal-like exponential dependence on $h$.
\end{abstract}

\maketitle

Percolation transitions are a cornerstone of statistical physics,
appearing across diverse fields, ranging from physics and chemistry
to, e.g., biology, ecology, and social
sciences~\cite{SA-book,Sahimi-book,Saberi-15}.  The standard
percolation theory was developed within the framework of lattice
models---more broadly, systems defined on graphs---in which sites or
bonds are occupied with some given probability $p$.  As $p$ is
increased, one reaches a threshold $p_c$ at which the occupied (or
activated) bonds form a percolating cluster, whose properties mirror
those of statistical systems undergoing continuous phase
transitions~\cite{Wilson-83,WK-74,Fisher-74,Wegner-76,PV-02}.  If the
activation probabilities of each site are independent, one recovers
the paradigmatic random percolation (RP) model, for which rigorous
exact results are available (see, e.g.,
Refs.~\cite{Kesten-book,Smirnov-01}), and which can be mapped onto the
$q$-state Potts models in the formal limit $q\to 1$~\cite{Wu-82},
allowing us to reinterpret the geometric transition as a standard one
in a spin system.  Percolation also emerges in dynamic protocols when
sites or bonds are reversibly added according to a stochastic
procedure~\cite{NZ-00}.  If the dynamic rules are local, the system
develops RP transitions as well.

In this paper we show that out-of-equilibrium percolation transitions
can emerge within dynamic, local, and reversible physical protocols,
in which a ferromagnetic system is driven across a magnetic
first-order transition.  We focus on the behavior of the paradigmatic
two-dimensional (2D) Ising model, when it is driven across its
low-temperature first-order transition line at zero magnetic field.
In particular, we analyze the out-of-equilibrium relaxational dynamics
after quenching the external magnetic field $h$ at fixed temperature.
Starting from negatively magnetized configurations, corresponding to
equilibrium states for $h<0$, the system evolves in the presence of a
positive fixed $h>0$, effectively crossing the magnetic first-order
transition line.  At a finite critical time $t_c$ after the quench, an
out-of-equilibrium transition occurs, which is marked by the
percolation of the largest clusters, and is related to the passage
from the metastable negatively magnetized phase to the stable positively
magnetized one. However, some critical features 
of this percolation transition differ from those the 
standard RP transition.

It is worth mentioning that non-RP percolation transitions have
already been obtained by devising particular nonlocal and irreversible
dynamic rules in some network models, such as the explosive
percolation (EP)
transitions~\cite{ASS-09,Ziff-09,CKPKK-09,RF-09,Ziff-10,CDGM-10,DM-10,
  RF-10,CD-11,NLT-11,GCBSP-11,CK-11,LKP-11,RW-11,RW-12,NTG-12,CDGM-14,
  BGMK-14,DN-15,Boetal-16,HS-18,LWD-23,YL-24}.  Note, however, that
the deviations from RP observed in EP transitions are essentially
related to the {\em ad hoc} nonlocal nature of the chosen dynamics.
On the other hand, in the percolation transitions considered here,
the dynamics is local, but the difference with RP stems from the
inherent out-of-equilibrium nature of the quenching process across 
first-order
transitions~\cite{Binder-76,Binder-87,Bray-94,RTMS-94,PV-17,PPV-18,
  Fontana-19,CCEMP-22, PV-24,PV-25,PRV-26}.  This shows that non-RP
transitions can also emerge in physical systems driven across
first-order transitions by physically realizable local dynamics.

We consider the 2D nearest-neighbor Ising model
\begin{equation}
  H(\beta,h) = - \beta \sum_{{\bm x},\mu} s_{\bm x} \, s_{{\bm x}+\hat{\mu}} -
  h \sum_{\bm x} s_{\bm x},\quad s_{\bm x}=\pm 1,
  \label{isiham}
\end{equation}
on a square lattice of size $L$ and $V=L^2$, with periodic
boundary conditions. The line $\beta > \beta_c = \ln (1 +
\sqrt{2})/2$ at $h=0$ is a first-order transition line, where the
infinite-volume magnetization $M = {\langle \, \sum_{\bm x} s_{\bm x}
  \, \rangle/V}$ is discontinuous, varying between $-M_0$ and $+M_0$
for $h=0^-$ and $h=0^+$ respectively, with $M_0=[1-(\sinh
  2\beta)^{-4}]^{1/8}$~\cite{ID-book}.

\begin{figure}[tbp]
  \includegraphics[width=0.9\columnwidth, clip]{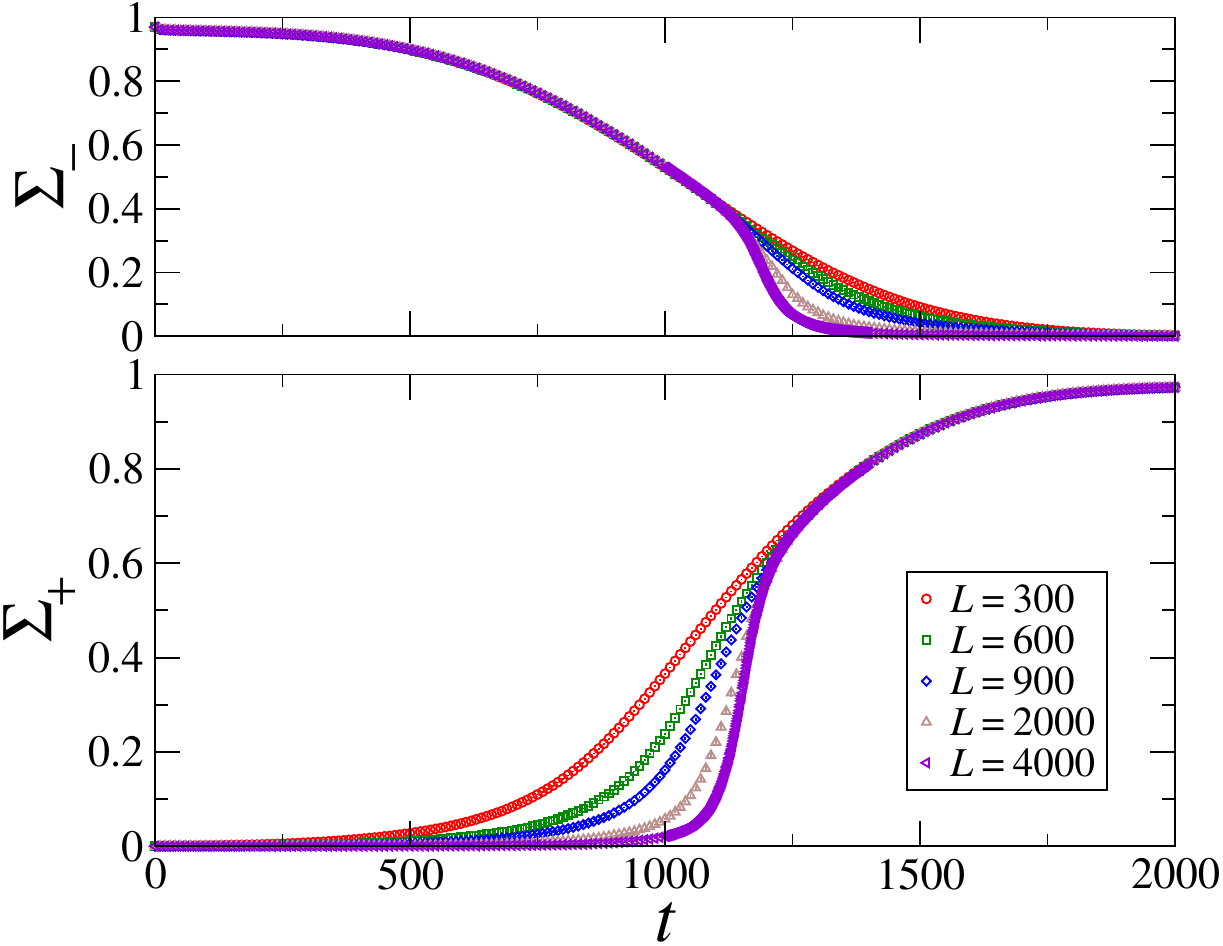}
  \caption{The rescaled average sizes $\Sigma_+$ (bottom) and
    $\Sigma_-$ (top) of the largest clusters of positive and negative
    magnetization, defined in Eq.~\eqref{drddef}, as a function of the
    post-quench time $t$, for $h=0.07$ and several sizes $L$.  Statistical
    errors are not visible, as they are smaller than the symbol size.}
\label{cluster2D}
\end{figure}

We analyze the evolution arising from a sudden quench of the magnetic
field across the $h=0$ first-order transition line, at fixed
$\beta>\beta_c$.  The
quenching protocol starts at $t=0$ from equilibrated
negatively magnetized configurations at $\beta>\beta_c$, effectively
corresponding to $h_i=0^-$. Then, the system evolves under a heat-bath
dynamics (a specific example of local relaxational
dynamics~\cite{Binder-76}) with $h>0$.  A time unit corresponds to a complete
lattice sweep. See App.~\ref{alprot}, for details. 
The evolution is monitored by the magnetization
\begin{equation}
  M(t) = {1\over V} \, \big\langle \sum_{\bm x} s_{\bm x} \big\rangle_t,
  \label{magn2}
\end{equation}
where the average is taken over the starting equilibrium
configurations and the post-quench trajectories at fixed time $t$.
The percolation nature of the dynamic transition is probed by looking
at clusters of equally oriented spins. For each configuration,
clusters with positive (negative) magnetization are the connected
components of the subset ${\cal B}_+$ (${\cal B}_-$) of lattice links
$\langle {\bm x}{\bm y}\rangle$ such that $s_{\bm x} = s_{\bm y} = 1$
($s_{\bm x} = s_{\bm y} = -1$).  They are determined using the
Hoshen-Kopelman algorithm~\cite{HK-76}. The size of each cluster is
defined as the number of sites belonging to it. We focus on the
fixed-time averages of the sizes $S_+$ and $S_-$ of the largest
clusters of positive and negative magnetization,
\begin{equation}
  \Sigma_\pm(t) = {\langle S_\pm \rangle_t\over V}, \qquad R_s(t)
  = {\langle S_- \rangle_t \over \langle S_+ \rangle_t}.
  \label{drddef}
\end{equation}
We report data at $\beta = 1.2\,\beta_c$ (results do not qualitatively
change for other $\beta$, as far as $\beta>\beta_c$) and some $h>0$.
For each $h$, we perform simulations for several sizes $L \le 4000$,
generating a large number of independent trajectories, 
between 2000 (for the largest systems) and 50000, which allow us to
compute $M(t)$ and $\Sigma_\pm(t)$.

\begin{figure}[tbp]
  \includegraphics[width=1.0\columnwidth, clip]{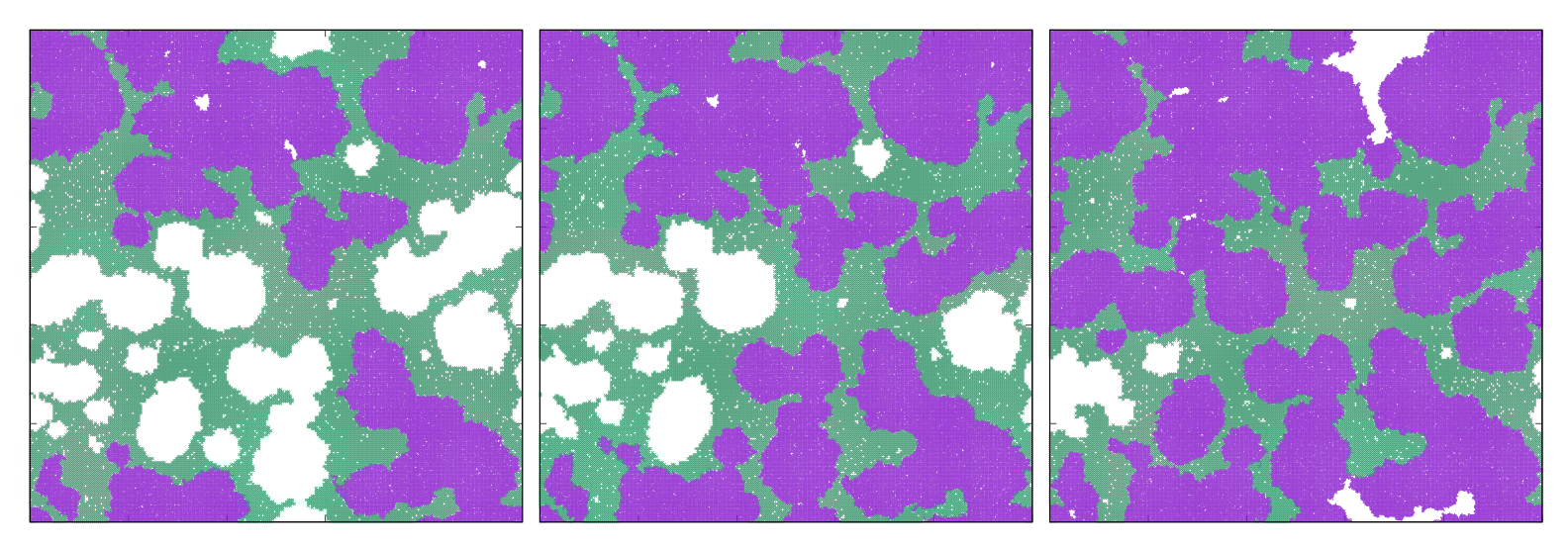}
  \caption{Snapshots of the configurations for $h = 0.07$ and
    $L=1000$, at times $t = 1144$ (left), $t = t_c = 1174$ (center),
    and $t = 1204$ (right). Violet and green sites belong to the
    largest clusters of positive and negative spins, respectively;
    white areas correspond to smaller (positive and negative)
    clusters. The size of the largest positive cluster is $368001$ for
    $t = 1144$, $498264$ for $t=t_c=1174$, and $637204$ for
    $t=1204$. Its growth is due to the merging with smaller clusters,
    since the boundaries of the clusters are essentially unchanged.}
\label{clusterfig2D}
\end{figure}

In Fig.~\ref{cluster2D} we report $\Sigma_\pm(t)$ for $h = 0.07$ and
several values of $L$.  We recognize two different regimes: a
small-$t$ regime where none of the positively magnetized clusters
fills a finite volume fraction of the lattice, while a fraction of the
volume belongs to a single negatively magnetized percolating cluster;
a large-$t$ regime where the opposite occurs, with a single
positively magnetized percolating cluster coexisting with small
negatively magnetized domains.  They are separated by a simultaneous
sharp variation of both $\Sigma_+(t)$ and $\Sigma_-(t)$, which becomes
sharper with increasing $L$, at a transition time $t_c
\approx 1200$.  As $t$ increases across $t_c$, positive-spin clusters
aggregate very rapidly, forming a single large cluster, while the
large negative-spin clusters disappear. Fig.~\ref{clusterfig2D} shows
a sketch of the typical rapid evolution of the configurations around
the transition time $t_c$.  We have obtained analogous outcomes for
other values of $h$, such as $h=0.06,\,0.05,\,0.045$.

\begin{figure}[tbp]
  \includegraphics[width=0.9\columnwidth, clip]{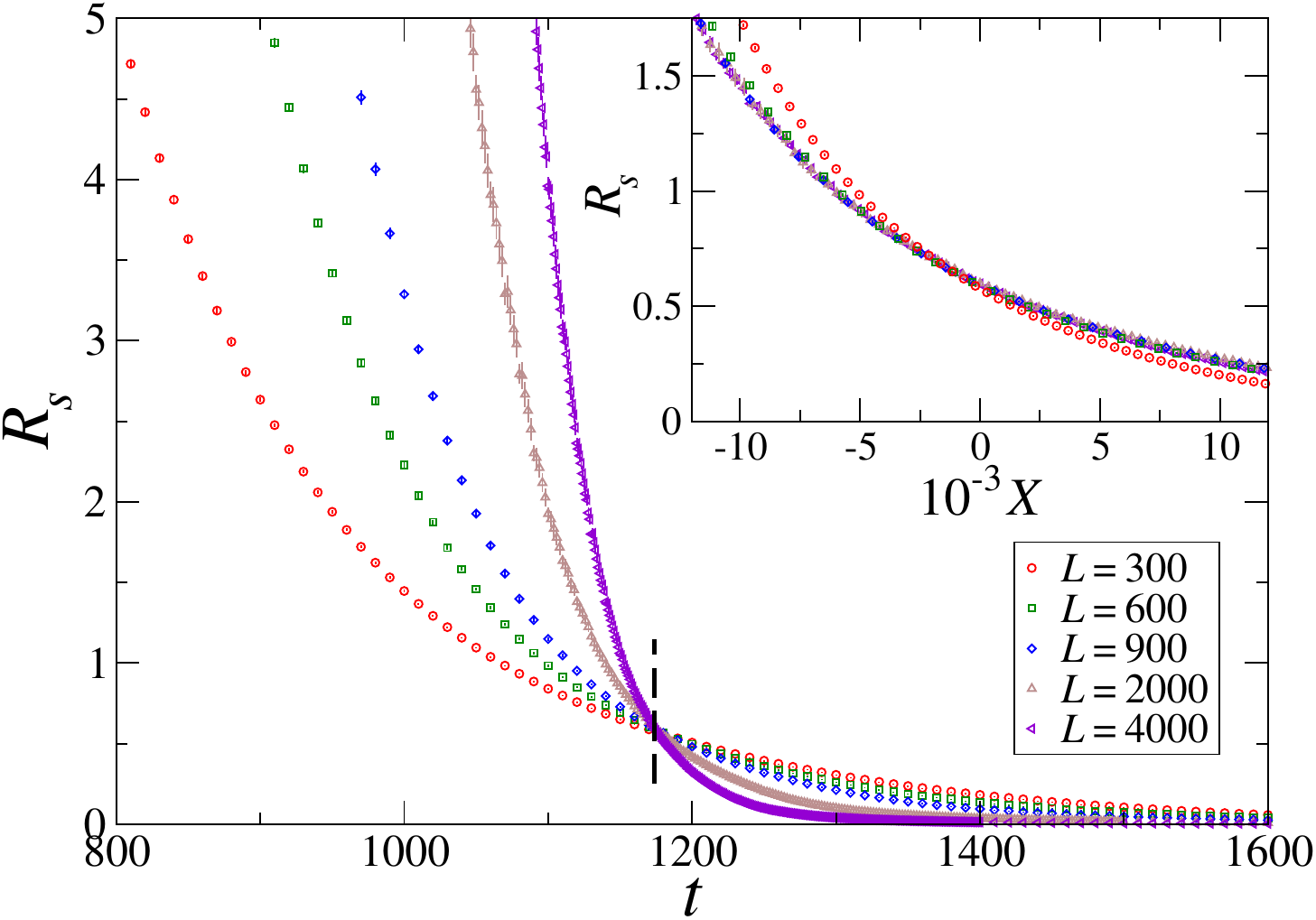}
  \caption{The time dependence of the ratio $R_s$ for $h = 0.07$. The
    vertical dashed line corresponds to the critical time $t_c =
    1174$.  The inset shows $R_s$ vs $X = (t-t_c) L^{w}$, with $w =
    0.68$.  The excellent collapse of the curves (in particular for $L
    > 600$) supports the Ansatz~\eqref{Ansatz-FSS}, confirming the
    equality $d_+=d_-$ of the fractal dimensions of positive and
    negative clusters.}
\label{clusterratioh0p07}
\end{figure}

These results suggest that the percolation of positive clusters and
inverse percolation of the negative clusters occurs at the same finite
time $t_c$ (this is likely a peculiar feature of 2D systems).  In this
case, $\Sigma_\pm(t)$ should have the typical finite-size scaling
(FSS) behavior at percolation
transitions~\cite{SA-book,Sahimi-book,Saberi-15},
\begin{equation}
  \Sigma_\pm(t,L) \approx L^{\delta_\pm} {\cal F}_\pm(X), \qquad
  X = (t-t_c) L^w,
  \label{Dpiu-scaling}
\end{equation}
where $\delta_\pm \equiv d-d_\pm = 2-d_\pm$ ($d$ is the dimension of
the system), $d_\pm$ are the critical fractal dimensions of the
clusters, and the exponent $w$ controls the approach to
criticality. The numerical results for the ratio $R_s$ also indicate
that $d_+ = d_-$. Indeed, as shown in Fig.~\ref{clusterratioh0p07},
they appear to scale as
\begin{equation}
  R_s(t,L) \equiv {\Sigma_-(t,L)\over \Sigma_+(t,L)} \approx
  {\cal F}_R(X).
  \label{Ansatz-FSS}
\end{equation}
Since $R_s\to \infty$ for $t<t_c$ and $R_s\to 0$ for $t>t_c$, the
above scaling behavior implies that curves for different $L$ cross at
a finite time, which converges to the critical time $t_c$ of the
percolation transition in the large-$L$ limit (resembling the behavior
of the Binder parameter in standard spin models~\cite{PV-02}). This is
clearly visible in Fig.~\ref{clusterratioh0p07}, where the $R_s$ data
for different $L$ cross at a finite time $t_c\approx 1174$.
Straightforward fits to the Ansatz~\eqref{Ansatz-FSS}, implemented
through low-order polynomial interpolations of ${\cal F_R}(X)$, lead
to the estimates $t_c = 1174(1)$ and $w = 0.680(15)$ (see
Table~\ref{table-td-2D}).  The inset of Fig.~\ref{clusterratioh0p07}
demonstrates the quality of the data collapse obtained using these
estimates. Analogous FSS behaviors can be found for other values of
$h$ (as an example, the scaling plot for $h=0.045$ is shown in
App.~\ref{infvol}). The corresponding estimates of $t_c$ and
$w$ are reported in Table~\ref{table-td-2D}.  The exponent $w$ depends
on $h$ and differs from the RP value $w_{\rm RP} =
3/4$~\cite{Kesten-book,Smirnov-01}. Moreover, it decreases as $h$ is
reduced, suggesting that it vanishes when $h\to 0$. As a consequence,
the time window $\Delta t$ over which the system changes phase, which
scales as $L^{-w}$, broadens as $h \to 0$. This behavior is consistent
with the progressive slowing down of the dynamics as $h$ decreases.

\begin{figure}[!b]
  \includegraphics[width=0.9\columnwidth, clip]{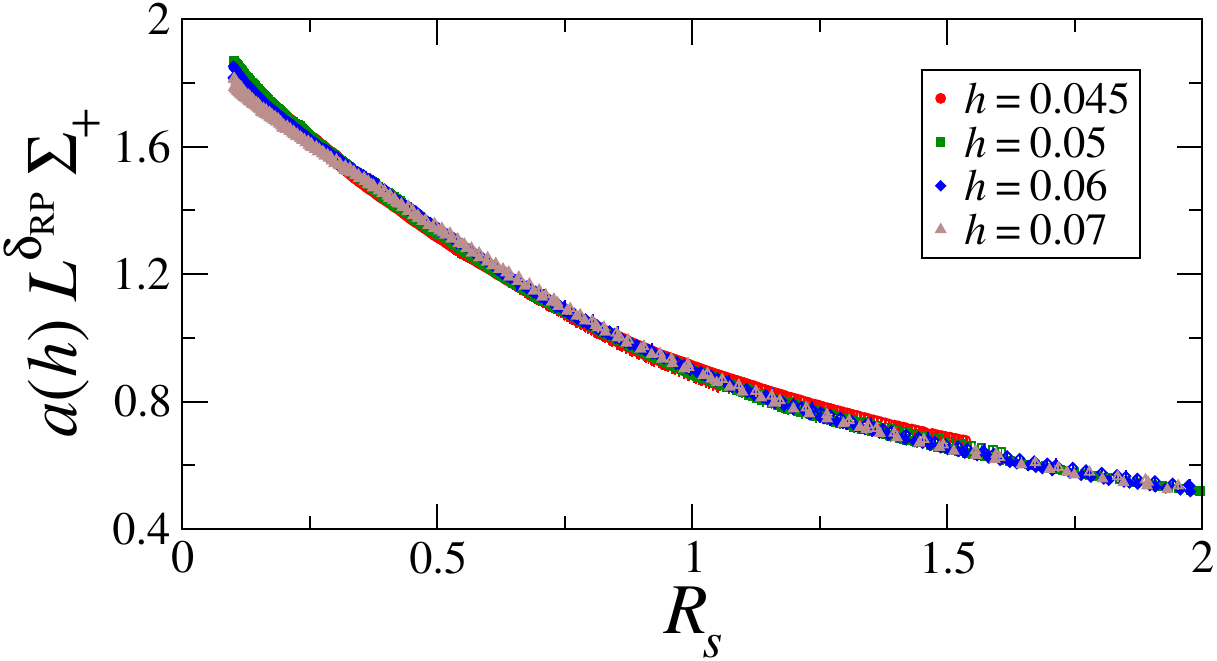}
  \caption{Plot of $a(h) L^{\delta_{\rm RP}} \Sigma_+$ vs $R_s$ for
    various $h$, where $a(h)$ is chosen to optimize the collapse of
    the data: Setting $a = 1$ for $h=0.045$, we use
    $a=1.137,1.096,1.037$ for $h=0.07,0.06,0.05$, respectively.  We
    report results for $L = 2000$ (empty symbols) and $L=4000$ (filled
    symbols).  The resulting scaling is good, with small deviations
    that can be explained by residual scaling corrections.}
\label{cluster2Dglobalscaling}
\end{figure}

The fractal-dimension exponent $\delta \equiv \delta_+=\delta_-$ can
be obtained by matching the data for $\Sigma_\pm(t)$ [see
  Eq.~\eqref{drddef}] with the asymptotic FSS behavior in
Eq.~\eqref{Dpiu-scaling}.  Since ${\cal F}_R(X)$ is monotonically
decreasing, we can express $X$ as a function of $R_s$ by inverting
Eq.~\eqref{Ansatz-FSS}, and write
\begin{equation}
  \Sigma_\pm(t,L) \approx L^{-\delta} \widetilde{\cal F}_\pm(R_s),
  \label{Dpiu-scaling2}
\end{equation}
in which only $\delta$ appears.  Standard fits to
Eqs.~\eqref{Dpiu-scaling} and~\eqref{Dpiu-scaling2} lead to the
estimates reported in Table~\ref{table-td-2D}, which vary little
around $\delta\approx 0.10$.  Notice that these estimates do not
account for possible systematic deviations due to scaling
corrections. Such corrections become more pronounced as $h$ decreases
and may induce deviations larger than the quoted errors.  Remarkably,
the numerical values of $\delta$ are very close to the RP exponent
$\delta_{\rm RP} = 5/48 \approx 0.104$~\cite{Kesten-book,SA-book}.
This suggests that the geometric properties of the critical clusters
may exhibit an $h$-independent universal behavior, and that the value
of $\delta$ coincides with $\delta_{\rm RP}$. To test this hypothesis,
we perform a consistency check. We fix $\delta = \delta_{\rm RP}$ for
all values of $h$ and we verify that the FSS function $\widetilde{\cal
  F}_\pm(R_s)$ obtained by $L^{\delta_{\rm RP}} \Sigma_\pm(t,L)$ is
universal up to a multiplicative $h$-dependent normalization factor.
As shown in Fig.~\ref{cluster2Dglobalscaling}, the resulting data
nicely collapse, confirming that the scaling properties are
$h$-independent and consistent with those of standard RP critical
clusters, even though the approach to the transition, most notably the
exponent $w$, retains a nontrivial dependence on $h$.

\begin{table}[t]
  \caption{Estimates of $t_c$, $w$, $\delta$, $\sigma_c = h (\ln
    t_c)^2$ and $\hat{\sigma}_c = (\sigma_c - \sigma_*) h^{-\theta}$
    [using $\theta = 0.64(4)$ and $\sigma_* = 3.03(3)$].  The errors
    take also into account the spread of the results when varying the
    minimum size allowed in the fit, the data interval considered, and
    the degree of the polynomial interpolation used for the scaling
    functions.}
  \label{table-td-2D}
  \vspace*{2mm}
  \begin{tabular}{lccccc}
    \hline\hline $h$ & $t_c$ & $w$ & $\delta$ &
    $\sigma_c$ & $\hat{\sigma}_c$
    \\ \hline
    0.07 & 1174(1) & 0.680(15) & 0.106(4) & 3.497(1) &2.56(4) \\
    0.06 & 1987(3) & 0.64(2)   & 0.106(10 & 3.460(1) &2.61(5) \\
    0.05 & 3904(3) & 0.56(2)   & 0.094(6) & 3.419(1) &2.65(6) \\
    0.045& 5921(3) & 0.52(2)   & 0.089(9) & 3.395(1) &2.66(7)
    \\ \hline\hline
  \end{tabular}
\end{table}

\begin{figure}[tbp]
  \includegraphics[width=0.9\columnwidth, clip]{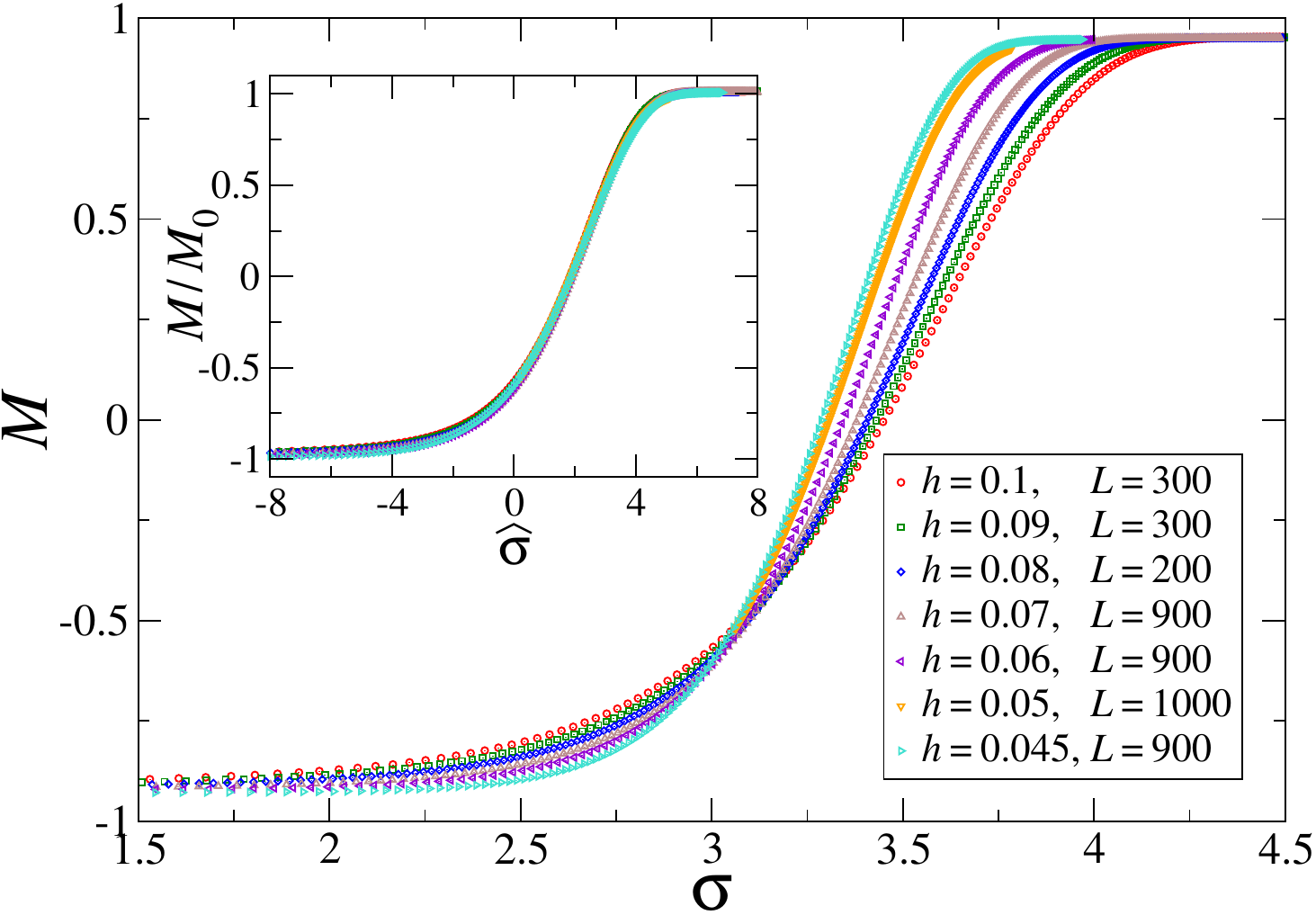}
  \caption{The magnetization $M(t)$ plotted vs $\sigma = h(\ln t)^2$.
    The size $L$ is large enough that the data can be regarded in their
    thermodynamic limit. The inset shows the ratio $M(t)/M_0$ vs
    $\hat\sigma=(\sigma-\sigma_*) h^{-\theta}$ with $\theta=0.64$ and
    $\sigma_*= 3.03$, where $M_0\approx 0.940259$ is the spontaneous
    magnetization for $\beta = 1.2\beta_c$.}
\label{resc}
\end{figure}

We now focus on the behavior of the magnetization $M(t)$ in the
thermodynamic limit (see App.~\ref{infvol}, for a
discussion of its determination from finite-size data). For finite
values of $h$, its post-quench evolution does not display finite-time
singularities.  However, a singular scaling behavior emerges for
$h\to 0$. Indeed, as discussed in App.~\ref{scalvol},
assuming that droplet nucleation is the relevant mechanism up to the
critical time where the phase change occurs, the infinite-volume
magnetization $M_\infty(t)$ is expected to scale as
\begin{equation}
  M_\infty(t,h) \approx {\cal M}_\infty(\sigma),\quad
  \sigma = h (\ln t)^{d/(d-1)} \!=\! h (\ln t)^2 .
  \label{sdef}
\end{equation}
This is confirmed by the numerical data. As shown in Fig.~\ref{resc},
the data of $M_\infty(t,h)$ corresponding to different values of $h$
get closer and closer as a function of $\sigma$ as $h$
decreases. Moreover, the curves cross at $\sigma_*\approx 3.0$. For
$\sigma$ close to $\sigma_*$, data show an additional scaling
behavior,
\begin{equation}
  M_\infty(t,h) \approx \widetilde{\cal M}_\infty(\hat{\sigma}), \quad
  \hat{\sigma} = (\sigma-\sigma_*) \, h^{-\theta},
  \label{estar}
\end{equation}
where $\theta>0$ (see the inset of Fig.~\ref{resc}).  Fits to
Eq.~\eqref{estar} yield $\sigma_*=3.03(3)$ and $\theta = 0.64(4)$.
The scaling form~\eqref{estar} implies that, in the limit $h\to 0$ and
for fixed $\sigma$, $M_\infty(t)$ is discontinuous at $\sigma_*$.
More precisely, as $h \to 0$, one finds $M_\infty(t)\to -M_0$ for
$\sigma < \sigma_*$, and $M_\infty(t) \to +M_0$ for $\sigma >
\sigma_*$, where $M_0$ denotes the spontaneous
magnetization~\cite{ID-book}.  The approach to these limiting values
is governed by corrections decreasing as $h^{\theta}$.

We now show that the time $t_*(h) = \exp{\sqrt{\sigma_*/h}}$,
corresponding to the crossing value $\sigma_*$, can be related to
the critical time $t_c(h)$. In
Table~\ref{table-td-2D} we report $\sigma_c = h (\ln t_c)^2$ and
$\hat{\sigma}_c = (\sigma_c - \sigma_*) h^{-\theta}$, which is
approximately constant within errors. This implies that $t_*(h) =
t_c(h) (1 - b h^\theta)$ with $b = \hat{\sigma}_c/(2 \sigma_*) =
0.45(2)$.  Therefore, for $h\to 0$ also $t_c$ diverges as
\begin{equation}
  t_c(h) = t_*(h) = \exp{\sqrt{\sigma_*/h}}.
  \label{tcsca}
\end{equation}
The exponent $\theta$ controls the approach to $h=0$. It is thus
natural to conjecture that also the $h$-dependence of $w$ is
controlled by $\theta$, thus $w\propto h^\theta$. Data are consistent:
Fits of $w(h)$ to $a h^\alpha$ give $\alpha = 0.60(2)$ (including all
values of $h$) and $\alpha = 0.72(8)$ (discarding the result for $h =
0.07$).

In conclusion, we report a study of the out-of-equilibrium
relaxational dynamics arising from quenches of the magnetic field
across the low-temperature first-order transition line of the 2D Ising
model, at fixed $\beta > \beta_c$ and for several values of the
post-quench field $h>0$.  Our results reveal the emergence of a
dynamic percolation transition at a finite post-quench critical time
$t_c$, which is marked by the simultaneous percolation of the largest
positive cluster and inverse percolation of the largest negative one.
For times $t$ close to $t_c$, the largest-cluster sizes show the
expected FSS behavior [see~Eq.~\eqref{Ansatz-FSS}], in terms of the
fractal dimensions $d_\pm$ and of a critical exponent $w$ controlling
the approach to criticality. We find that the fractal dimensions
$d_\pm$ are equal, apparently independent of $h$, and consistent with
the RP value $d_{\rm RP} = 91/48$. On the other hand, the 
exponent $w$, controlling the approach to criticality, retains a
nontrivial dependence on the post-quench $h$, and differs from the RP
value $w_{\rm RP}=3/4$.  We find $w<w_{\rm RP}$ for all considered
values of $h$, with $w$ decreasing as $h$ is reduced, apparently
vanishing as $w \sim h^\theta$.  From a renormalization-group point of
view, this behavior may be interpreted as a line of $h$-dependent
fixed points sharing the same metric (magnetic, in the equivalent spin
representation) critical exponents, but differing in the scaling
behavior governing the dynamical approach to criticality.

The critical percolation time $t_c$ is related to the typical time
$t_*$ at which the system switches from the metastable negatively
magnetized phase to the positively magnetized one, and therefore to
the spinodal-like behavior of the magnetization in the small-$h$
limit. This implies that the percolation time $t_c$ exhibits a
spinodal-like exponential dependence on $h$, cf.  Eq.~\eqref{tcsca},
somehow mimicking the decay of a false vacuum. Our results suggest
that the mechanism driving the transition from the metastable to the
stable phase is not the independent growth of isolated droplets, but
rather their collective aggregation dynamics, whereby droplets
predominantly grow through merging processes (see 
App.~\ref{scalvol}).

The percolative behavior outlined in this work is most likely generic.
Indeed, percolation transitions are also expected under alternative
quenching protocols---such as slowly varying $h$ across the transition
point $h=0$ (see, e.g., Refs.~\cite{PV-17,PV-24,PV-25,PRV-26})--- and
in higher dimensions. However, in three-dimensional Ising systems the
pattern turns out to be quite different~\cite{BPV-26}: positive and
negative clusters undergo separate percolation transitions, with an
intermediate dynamical phase in which percolating clusters of both
signs coexist.  More generally, percolation transitions are expected
in systems with discrete symmetries (systems with continuous
symmetries may behave differently, due to the presence of
long-distance Goldstone-like modes~\cite{PV-16}), with features that
may depend on the dimensionality and the number of low-temperature
phases.  Percolation may also be relevant at thermal first-order
transitions, such as those occurring in $q$-state Potts models for
large $q$ (some indications in this direction can already be found in
the results reported in Ref.~\cite{PV-17}). These issues call for
further investigation.

We finally remark that the out-of-equilibrium percolation transitions
across magnetic first-order transitions represent a new type of non-RP
percolative behavior, which naturally emerges in physical systems
governed by physically realizable local dynamics. This should be
contrasted with other examples of non-RP percolation transitions
arising from nonlocal and irreversible {\em ad hoc} dynamical rules,
as in certain network-growth models exhibiting EP
transitions~\cite{ASS-09,Ziff-09,CKPKK-09,RF-09,Ziff-10,CDGM-10,DM-10,
  RF-10,CD-11,NLT-11,GCBSP-11,CK-11,LKP-11,RW-11,RW-12,NTG-12,CDGM-14,
  BGMK-14,DN-15,Boetal-16,HS-18,LWD-23,YL-24}.

\appendix

\section{}
\label{endmatter}

Here we report some additional details on the dynamics and quenching
protocol, and on the determination of the infinite-volume magnetization
$M_\infty(t)$. Moreover, we summarize the arguments leading to the
definition of the scaling variable $\sigma = h (\ln t)^2$
[cf.~Eq.~\eqref{sdef}].

\subsection{Dynamics  and quenching protocol}
\label{alprot}

We analyze the out-equilibrium evolution arising from a sudden quench
of the magnetic field across the low-temperature first-order
transition line, at fixed $\beta>\beta_c$. We consider the single-spin
heat-bath dynamics, an example of a purely relaxational
dynamics~\cite{Binder-76}: at each site, a new spin is chosen using
the conditional probability distribution $P \sim e^{-H(\beta,h)}$ at
fixed neighboring spins.  Spins are updated using a checkerboard
scheme: we first update all spins at even sites, then all spins at odd
sites.  A time unit corresponds to a complete lattice sweep. The
resulting dynamics is not strictly reversible, as the two sublattices
are updated sequentially.  This also implies that the time $t$ can
only take integer values.  A reversible dynamics could be obtained by
randomly choosing spins at each step. This choice, giving rise to a
slightly slower dynamics (the typical time scales differ by a factor
of approximately four~\cite{PV-17b}), would also allow for noninteger
values of the time variable.  Indeed, if the time unit corresponds to
$L^2$ spin updates, so as to obtain a well-defined infinite-volume
limit at fixed $t$, one might consider time steps of order $1/L^2$. In
the $L\to \infty$ limit, this naturally leads to a continuous-time
evolution.

The quenching protocol starts at $t=0$ from equilibrated negatively
magnetized configurations at $\beta>\beta_c$ and $h=0$. The volume is
large enough, so we are effectively considering $h_i=0^-$ in the
thermodynamic limit.  Indeed, for $h=0$, the typical time needed to
make a transition from the negative phase to the opposite one scales
as $L^2 \exp(2 \beta \kappa L)$, where $\kappa$ is the planar
interface tension (see, e.g., Ref.~\cite{PV-17b}). In particular, for
$\beta=1.2\beta_c$, it scales as $L^2 \exp(0.666 L)$, corresponding to
$3 \times 10^{68}$ for $L=200$. Since we only consider lattices with
$L\ge 200$, transitions never occur and therefore the starting
configurations have been simply obtained by performing a heat-bath
Monte Carlo simulation at $h=0$, starting from a configuration with
$s_{\bm x} = -1$ for all values of $\bm x$.  We note that 
starting from configurations equilibrated at negative values of $h$
would not change the general picture and would only lead to a shift of 
the critical time~$t_c$.

\subsection{Additional numerical details}
\label{infvol}

\begin{figure}[tbp]
  \includegraphics[width=0.9\columnwidth, clip]{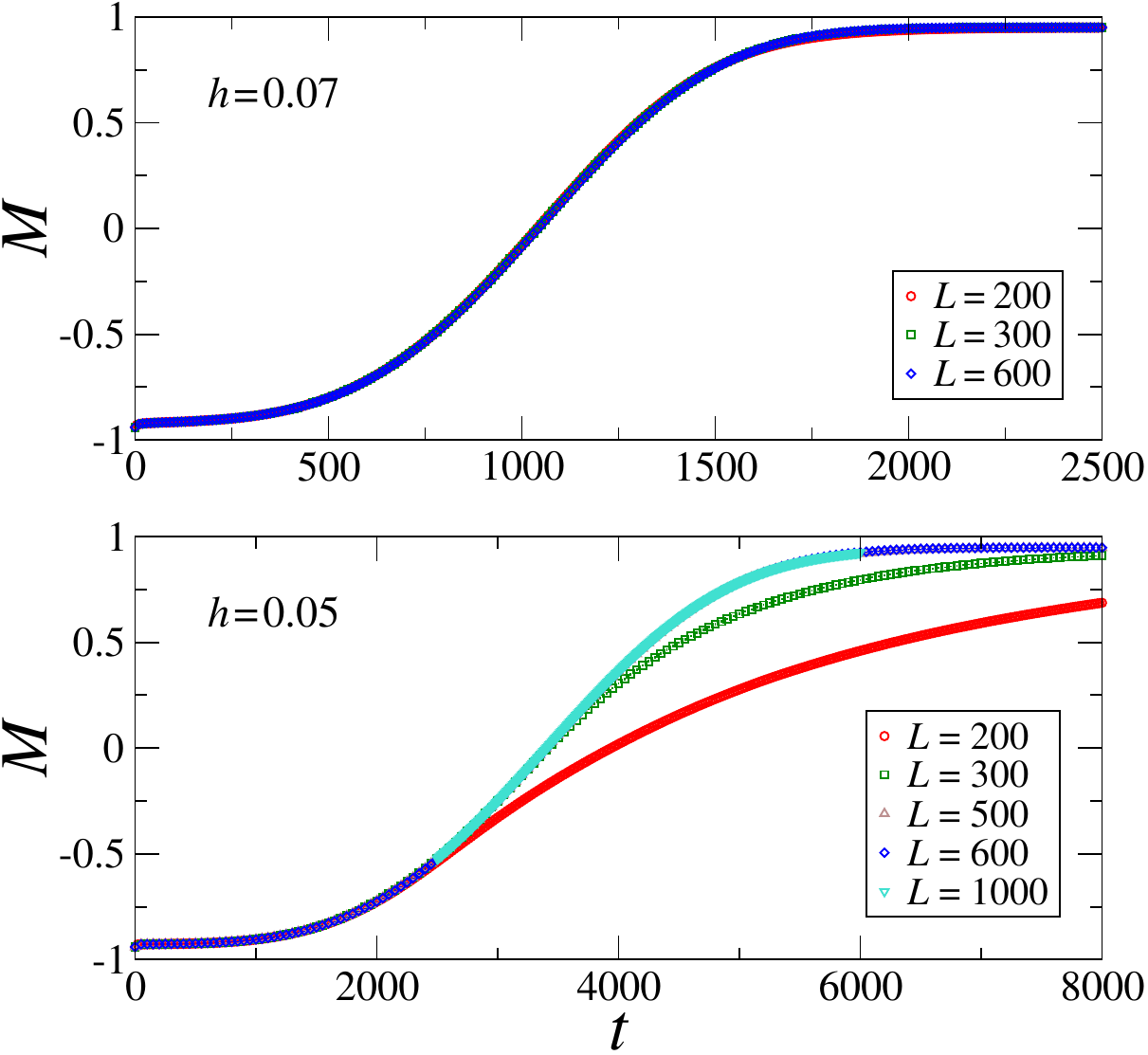}
  \caption{Time evolution of the magnetization $M(t)$ for $h=0.07$
    (top) and $h = 0.05$ (bottom) and several sizes $L$.  Statistical
    errors are hardly visible (for $L=600$, the errors are at most
    $0.003$ and $0.004$ for $h=0.07$ and $h=0.05$, respectively).
    Data clearly converge to a limiting curve as $L$ increases, which
    provides a good approximation of $M_\infty(t)$. }
\label{rawlts-suppl}
\end{figure}

\begin{figure}[tbp]
  \includegraphics[width=0.9\columnwidth, clip]{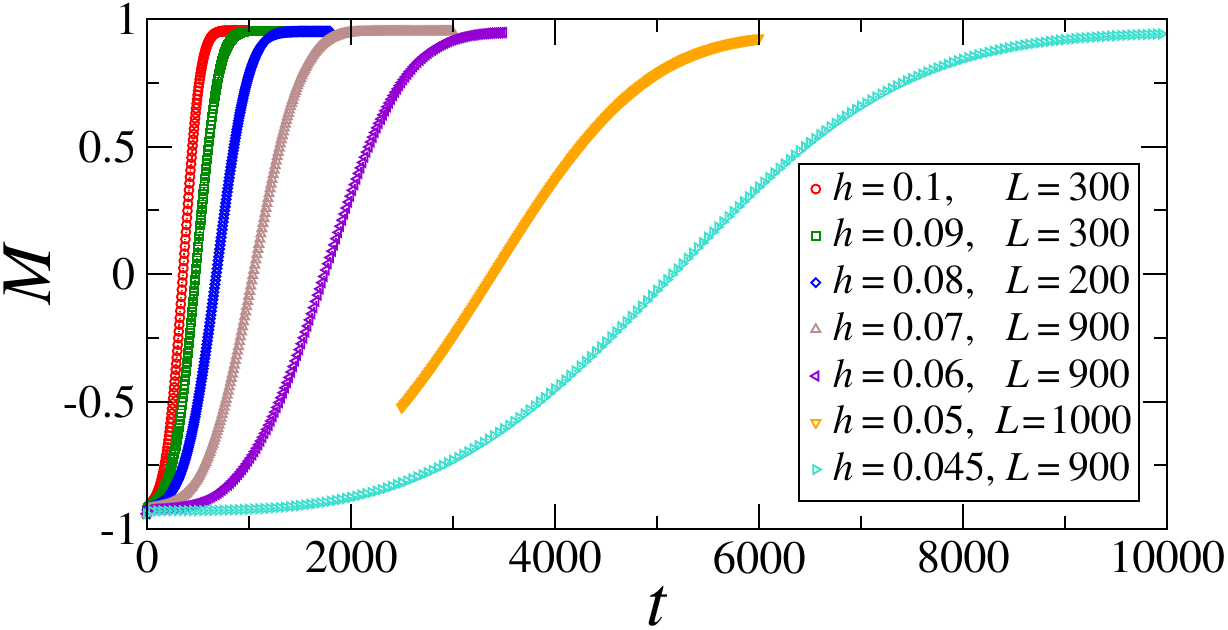}
  \caption{Time dependence of $M(t)$ for several values of $h$, and
    sufficiently large $L$ to provide their infinite-volume limit
    $M_\infty(t)$ (statistical errors are hardly visible).}
\label{magn-allh-suppl}
\end{figure}

In our numerical study we discuss the scaling behavior of the
infinite-volume magnetization $M_\infty(t)$ at fixed $h$.  To
determine it for a given value of $h$, we increase $L$ until the
average magnetization curves become approximately $L$-independent. The
curves corresponding to the largest $L$ then provide the infinite-size
$M_\infty(t)$ for the given values of $h$.  As an example, in
Fig.~\ref{rawlts-suppl} we show the magnetization data for two values
of $h$ and different lattice sizes.  For $h = 0.07$, simulations on
lattices of size $L=200$ already provide the infinite-volume
magnetization curves. As $h$ decreases, finite-size effects become
larger. For example, for $h=0.05$ (see Fig.~\ref{rawlts-suppl}), the
large-$L$ limit (within errors) is obtained for $L\gtrsim 600$
($L\gtrsim 900$ for $h=0.045$).  The infinite-volume results for
different values of $h$ are reported in Fig.~\ref{magn-allh-suppl} for
values of $h$ in the interval $[0.045,0.10]$.  For $h = 0.10$, the
transition to the positively magnetized phase occurs rapidly and
indeed, for larger values of $h$, it is difficult to identify an
out-of-equilibrium regime. As $h$ decreases, the system persists in
the negatively magnetized state for an increasingly long time
interval.

\begin{figure}[tbp]
  \includegraphics[width=0.9\columnwidth, clip]{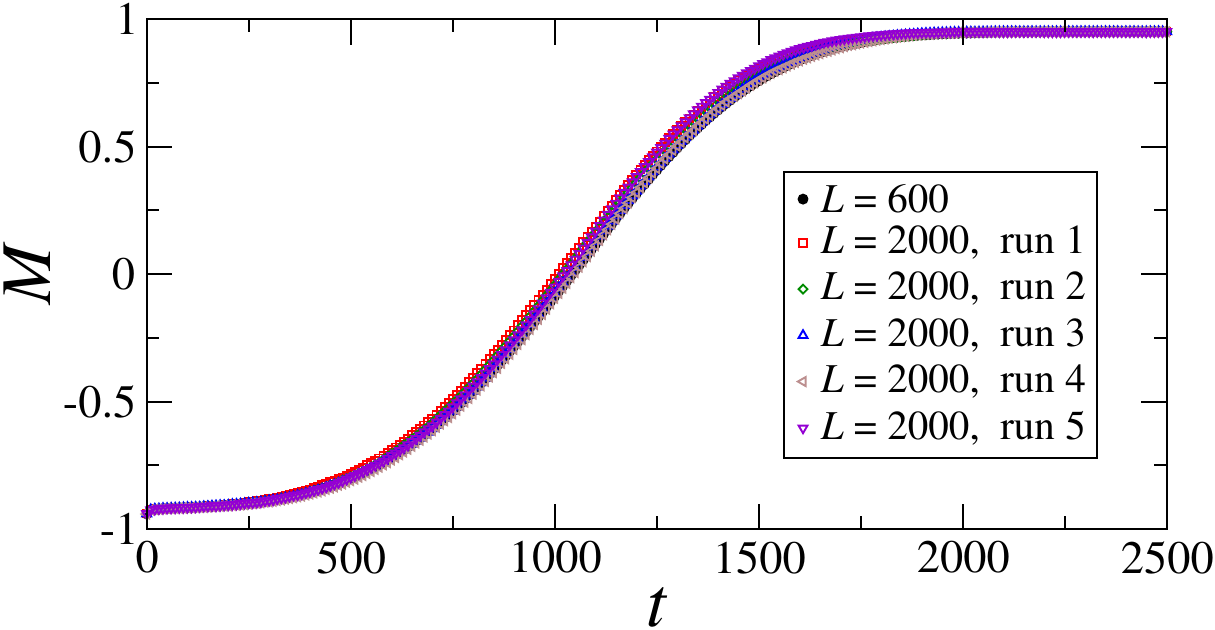}
  \caption{Comparison of the average magnetization $M(t)$ for $L=600$
    with the instantaneous magnetization for five different runs on a
    large lattice with $L=2000$, for $h=0.07$.}
  \label{magn-singoli-suppl}
\end{figure}

It is important to note that the infinite-volume magnetization curves
do not only provide information on the average behavior of the system
as a function of $t$, but represent the evolution of the magnetization
of any large system.  Indeed, self-averaging holds, so that fluctuations
in the evolution vanish for $L\to \infty$. This is evident from
Fig.~\ref{magn-singoli-suppl}, where we compare the evolution of the
instantaneous magnetization on large systems (here $L=2000$) with the
average behavior computed on smaller lattices.

\begin{figure}[tbp]
  \includegraphics[width=0.9\columnwidth, clip]{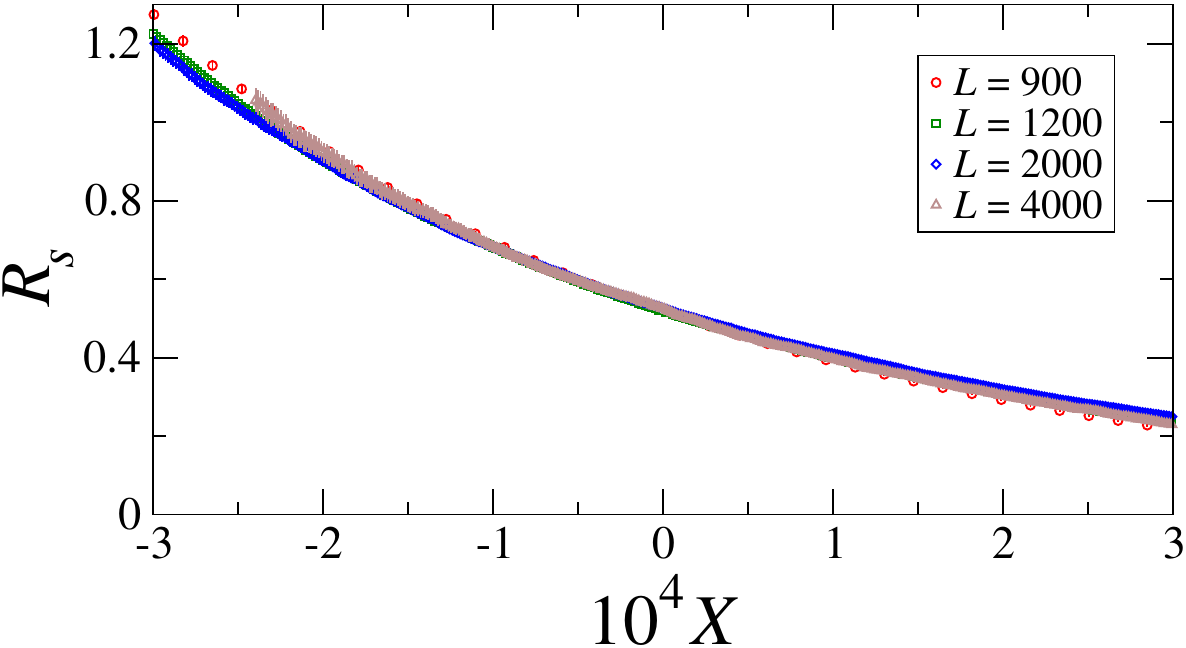}
  \caption{The ratio $R_s$ vs $X = (t-t_c) L^{w}$ for $h = 0.045$,
    with $t_c = 5921$ and $w = 0.52$.}
  \label{cluster2Dscalingh0p045-suppl}
\end{figure}

Finally, we report some additional data to support the scaling
relation (\ref{Ansatz-FSS}).  Fig.~\ref{cluster2Dscalingh0p045-suppl}
shows the ratio $R_s$ vs $X=(t-t_c) L^w$ for $h=0.045$. It is the
analogue of Fig.~\ref{clusterratioh0p07}, where we report results for
$h=0.07$.  Also in this case, data scale quite well.

\subsection{Scaling behavior from droplet nucleation}
\label{scalvol}

We outline here the arguments leading to the scaling
Ansatz~\eqref{sdef} for the magnetization in 2D systems. We assume
that, for the infinite-volume dynamics across the first-order
transition line, the relevant scaling variable is the magnetic energy
of droplets of positive spins nucleated in the dynamical process,
which are eventually responsible for the transition from the
metastable negatively magnetized state to the positively magnetized
one.  Under this hypothesis, the relevant scaling variable is $\sigma
\sim h R(t)^d$ ($d$ is the spatial dimension), where $R(t)$ denotes
the typical droplet size at time $t$. Assuming droplets to have smooth
boundaries, the time needed to create a droplet of size $R$ should
scale as $\exp(c R^{d-1})$ in $d$ dimensions, so $R(t) \sim (\ln
t)^{1/(d-1)}$.  Therefore we obtain Eq.~\eqref{sdef} for $d=2$.

As discussed in the main text, the scaling of the data in terms of
$\sigma$ implies that for small $h$ the phase change of 2D systems
occurs on time scales
\begin{equation}
  t_* \sim \exp\sqrt{\sigma_*/h}.
  \label{tstar-suppl}
\end{equation}
Note that $t_*$ is significantly shorter than the time
scale obtained by requiring isolated droplets to be energetically
stable. Indeed, the energy of a positive-spin droplet surrounded by
negative spins is
\begin{equation}
  E = - a h R^d + b R^{d-1}, \qquad a,b>0.
  \label{droene}
\end{equation}
It is immediate to verify that the energy decreases with increasing
$R$, only if $R$ is larger than a threshold $R_c$ that scales as
$1/h$. Droplets of size $R_c$ require a time of order $\exp( c R_c)
\sim \exp(\bar{c}/h)$ to be generated~\cite{RTMS-94}.  For $h\to 0$,
this time scale is significantly larger than the one defined in
Eq.~\eqref{tstar-suppl}, indicating that in two dimensions
 the transition occurs when
typical isolated droplets are too small to be energetically
stable. Thus, the main mechanism responsible for the transition in 2D
Ising systems is not the independent growth of isolated droplets, but
rather their aggregation dynamics: droplets increase their size
predominantly by merging, a dynamical process that is always
energetically favored.  This mechanism allows the system to generate
stable positive-spin domains on time scales much shorter than
$\exp(\bar{c}/h)$ and is responsible for the percolation transition
that marks the passage from one phase to the other.

The relevance of the merging dynamics in the critical region is
particularly evident in Fig.~\ref{clusterfig2D}, where we report the
configurations of the system at three different times in the critical
region: we consider $t=t_c-30,t_c,t_c+30$, with $t_c = 1174$ (results
are for $h=0.07$). In this short time interval, the cluster boundaries
barely move and the large increase in the size of the largest positive
cluster ($\Sigma_+$ approximately takes the values 0.37, 0.50,
and 0.64, as time increases) is due to the merging of the largest
cluster with smaller positive-spin clusters.


\begin{thebibliography}{99}

\bibitem{SA-book} D. Stauffer and A. Aharony, {\em Introduction to
  percolation theory} (Taylor and Francis, London, 1994).

\bibitem{Sahimi-book} M. Sahimi, {\em Applications of percolation
  theory} (CRC Press, 1994).
  
\bibitem{Saberi-15} A. A. Saberi, Recent advances in percolation
  theory and its applications, Phys. Rep. {\bf 578}, 1 (2015).

\bibitem{Wilson-83} K. G. Wilson, The renormalization group and
  critical phenomena, Rev. Mod. Phys. {\bf 55}, 483 (1983).
  
\bibitem{WK-74} K. G. Wilson and J. B. Kogut, The renormalization
  group and the epsilon expansion, Phys. Rep. {\bf 12}, 75 (1974).

\bibitem{Fisher-74} M. E. Fisher, The renormalization group in the
  theory of critical behavior, Rev. Mod. Phys. {\bf 46}, 597 (1974);
  Erratum: Rev. Mod. Phys. {\bf 47}, 543 (1975).

\bibitem{Wegner-76} F. J. Wegner, The Critical State, General Aspects,
  C. Domb, C. and M. S. Green editors, {\em Phase transitions and
    critical phenomena}, vol. 6, (Acad. Press, London, 1976).

\bibitem{PV-02} A. Pelissetto and E. Vicari, Critical phenomena and
  renormalization group theory, Phys. Rep. {\bf 368}, 549 (2002).

\bibitem{Kesten-book}
  H. Kesten, {\em Percolation Theory for Mathematicians},
  (Birkh\" auser, 1982).

\bibitem{Smirnov-01} S. Smirnov, Critical percolation in the plane:
  Conformal invariance, Cardy's formula, scaling limits, C. R.
  Acad. Sci. (Paris) I {\bf 333}, 239 (2001).

\bibitem{Wu-82} F. Y. Wu, The Potts model, Rev. Mod. Phys. {\bf 64},
  235 (1982).
  
\bibitem{NZ-00} M. E. J. Newman and R. M. Ziff, Efficient Monte Carlo
  algorithm and high-precision results for percolation,
  Phys. Rev. Lett. {\bf 85}, 4104 (2000); A fast Monte Carlo algorithm
  for site or bond percolation, Phys. Rev. E {\bf 64}, 016706 (2001).

\bibitem{ASS-09} D. Achlioptas, R. M. D'Souza, and J. Spencer,
  Explosive percolation in random networks, Science {\bf 323}, 1453
  (2009).

\bibitem{Ziff-09} R. M. Ziff, Explosive growth in biased dynamic
  percolation on two-dimensional regular lattice networks,
  Phys. Rev. Lett. {\bf 103}, 045701 (2009).

\bibitem{CKPKK-09} Y. S. Cho, J. S. Kim, J. Park, B. Kahng, and
  D. Kim, Percolation transitions in scale-free networks under the
  Achlioptas process, Phys. Rev. Lett. {\bf 103}, 135702 (2009).
  
\bibitem{RF-09} F. Radicchi and A. Fortunato, Explosive percolation
  in scale-free networks Phys. Rev. Lett. {\bf 103}, 168701 (2009).

\bibitem{Ziff-10} R. M. Ziff, Scaling behavior of explosive
  percolation on the square lattice, Phys. Rev. E {\bf 82}, 051105
  (2010).

\bibitem{CDGM-10}
  R. A. Costa, S. N. Dorogovtsev, A. V. Goltsev, and J. F. F. Mendes,  
  Explosive percolation transition is actually continuous,
  Phys. Rev. Lett. {\bf 105}, 255701 (2010).

\bibitem{DM-10} R. M. D'Souza and M. Mitzenmacher, Local cluster
  aggregation models of explosive percolation, Phys. Rev. Lett.
  {\bf 104}, 195702 (2010).

\bibitem{RF-10}  
  F. Radicchi and S. Fortunato, Explosive percolation:
  A numerical analysis, Phys. Rev. E {\bf 81}, 036110 (2010).  

\bibitem{CD-11}
  W. Chen and R. M. D'Souza, 
  Explosive percolation with multiple giant components,
  Phys. Rev. Lett. {\bf 106}, 115701 (2011).

\bibitem{NLT-11} J. Nagler, A. Levina, and M. Timme, Impact of single
  links in competitive percolation, Nat. Phys. {\bf 7}, 265 (2011)

\bibitem{GCBSP-11} P. Grassberger, C. Christensen, G. Bizhani,
  S.-W. Son, and M. Paczuski, Explosive percolation is continuous, but
  with unusual finite size behavior, Phys. Rev. Lett. {\bf 106},
  225701 (2011).

\bibitem{CK-11} Y. S. Cho and B. Kahng, Suppression effect on
  explosive percolation, Phys. Rev. Lett. {\bf 107}, 275703 (2011).

\bibitem{LKP-11} H. K. Lee, B. J. Kim, and H. Park, Continuity of the
  explosive percolation transition, Phys. Rev. E {\bf 84}, 020101(R)
  (2011).

\bibitem{RW-11}  
  O. Riordan and L. Warnke, Explosive percolation is continuous,
  Science {\bf 333}, 322 (2011).
  
\bibitem{RW-12}  
  O. Riordan and L. Warnke, Achlioptas processes are not
  always self-averaging, Phys. Rev. E {\bf 86}, 011129 (2012).
  
\bibitem{NTG-12}
  J. Nagler, T. Tiessen, and H. W. Gutch,
  Continuous percolation with discontinuities,
  Phys. Rev. X {\bf 2}, 031009 (2012).

\bibitem{CDGM-14}
  R. A. da Costa, S. N. Dorogovtsev, A. V. Goltsev, and J. F. F.
  Mendes, Critical exponents of the explosive percolation 
  transition, Phys. Rev. E {\bf 89}, 042148 (2014).

\bibitem{BGMK-14}  
  N. Bastas, P. Giazitzidis, M. Maragakis, and K. Kosmidis, Explosive
  percolation: Unusual transitions of a simple model, Physica A
  {\bf 407}, 54 (2014).

\bibitem{DN-15}
  R. M. D'Souza and J. Nagler,
  Explosive percolation: Novel critical and supercritical phenomena,  
  Nat. Phys. {\bf 11}, 531 (2015).

\bibitem{Boetal-16}
  S. Boccaletti, J. A. Almendral, S. Guan, I. Leyva, Z. Liu,
  I. Sendi\~na-Nadal, Z. Wang, and Y. Zou, Explosive transitions
  in complex networks structure and dynamics: Percolation and
  synchronization, Phys. Rep. {\bf 660}, 1 (2016).

\bibitem{HS-18}
  M. K. Hassan and M. M. H. Sabbir,  
  Product-sum universality and Rushbrooke inequality in explosive percolation,
  Phys. Rev. E {\bf 97}, 050102 (2018).

\bibitem{LWD-23} M. Li, J, Wang, and Y. Deng, Explosive percolation
  obeys standard finite-size scaling in an event-based ensemble,
  Phys. Rev. Lett. {\bf 130} 147101 (2023).

\bibitem{YL-24}
  L. Yang and M. Li, Emergence of biconnected clusters in explosive 
  percolation,  Phys. Rev. E {\bf 110}, 014122 (2024).

\bibitem{Binder-76} K. Binder, Monte Carlo investigations of phase
  transitions and critical phenomena.  {\em Phase Transitions and
    Critical Phenomena}, Domb, C. \& Green, M. S. (eds.)  vol. 5b, 1
  (Academic Press, London, 1976).  

\bibitem{Binder-87} K. Binder, Theory of first-order phase
  transitions, Rep. Prog. Phys. {\bf 50}, 783 (1987).  

\bibitem{Bray-94} A. J. Bray, Theory of phase-ordering kinetics,
  Adv. Phys. {\bf 43}, 357 (1994).

\bibitem{RTMS-94} P. A. Rikvold, H. Tomita, S. Miyashita, and
  S. W. Sides, Metastable lifetimes in a kinetic Ising model:
  Dependence on field and system size, Phys. Rev. E {\bf 49}, 5080
  (1994).

\bibitem{PV-17} A. Pelissetto and E. Vicari, Dynamic off-equilibrium
  transition in systems slowly driven across thermal first-order
  transitions, Phys. Rev. Lett. {\bf 118}, 030602 (2017).

\bibitem{PPV-18} H. Panagopoulos, A. Pelissetto, and E. Vicari, Dynamic
  scaling behavior at thermal first-order transitions in systems with
  disordered boundary conditions, Phys. Rev. D {\bf 98}, 074507 (2018).

\bibitem{Fontana-19} P. Fontana, Scaling behavior of Ising systems at
  first-order transitions, J. Stat. Mech. 063206 (2019).  
  
\bibitem{CCEMP-22} F. Corberi, L. F. Cugliandolo, M. Esposito,
  O. Mazzarisi, and M. Picco, How many phases nucleate in the
  bidimensional Potts model?, J. Stat. Mech. 073204 (2022).

\bibitem{PV-24} A. Pelissetto and E. Vicari, Scaling behaviors at quantum
  and classical first-order transitions, in {\em Fifty Years of the
    Renormalization Group}, Chap. 27, dedicated to the memory of
  Michael E. Fisher, edited by A. Aharony, O. Entin-Wohlman, D. Huse,
  and L. Radzihovsky, World Scientific (2024) [arXiv:2302.08238]

\bibitem{PV-25} A. Pelissetto and E. Vicari, Out-of-equilibrium
  spinodal-like scaling behaviors across the magnetic first-order
  transitions of two-dimensional and three-dimensional Ising systems,
  Phys. Rev. E {\bf 113}, 014107 (2026).

\bibitem{PRV-26} A. Pelissetto, D. Rossini, and E. Vicari,
  Out-of-equilibrium spinodal-like scaling behaviors at the thermal
  first-order transitions of three-dimensional $q$-state Potts models,
  Phys. Rev. E {\bf 113}, 024137 (2026).

\bibitem{ID-book}  
  C. Itzykson and J.-M. Drouffe, Statistical Field Theory: Volume 1.
  From Brownian Motion to Renormalization and Lattice
  Gauge Theory (Cambridge University Press, Cambridge, UK, 1989).
  
\bibitem{HK-76}
  J. Hoshen and R. Kopelman, Percolation and cluster distribution. I.
  Cluster multiple labeling technique and critical concentration
  algorithm, Phys. Rev. B {\bf 14}, 3438 (1976).

\bibitem{BPV-26} A. Pelissetto, D. Rossini, and E. Vicari, in
  preparation.

\bibitem{PV-16} A. Pelissetto and E. Vicari, Off-equilibrium scaling
  behaviors driven by time-dependent external fields in
  three-dimensional O($N$) vector models, Phys. Rev. E {\bf 93},
  032141 (2016).

\bibitem{PV-17b} A. Pelissetto and E. Vicari, Dynamic finite-size
  scaling at first-order transitions, Phys. Rev. E {\bf 96}, 012125
  (2017).
  
\end{thebibliography}
\end{document}